\documentclass[preprint]{aastex}

\shorttitle{White Dwarfs in NGC 2477}
\shortauthors{Jeffery, et al.}

\begin{document}

\title{The White Dwarf Age of NGC 2477\footnote{Based on observations made 
with the NASA/ESA \textit{Hubble Space Telescope}, obtained at the Space 
Telescope Science Institute, which is operated by the Association of 
Universities for Research in Astronomy, Incorporated, under NASA contract 
NAS5-26555.}}

\author{Elizabeth J. Jeffery}
  \affil{Space Telescope Science Institute, 3700 San Martin Drive,
    Baltimore, MD 21218, USA}
\email{jeffery@stsci.edu}
\author{Ted von Hippel}
  \affil{Department of Physics and Astronomy, Siena College, Loudonville, NY, USA}
\author{Steven DeGennaro}
  \affil{Department of Astronomy, University of Texas at Austin,
    Austin, TX, USA}
\author{David A. van Dyk}
  \affil{Department of Statistics, The University of California,
    Irvine, CA, USA}
\author{Nathan Stein}
  \affil{Department of Statistics, Harvard University,
    Cambridge, MA, USA}
\author{William H. Jefferys}
  \affil{Department of Astronomy, The University of Texas at Austin, Austin, TX, USA}
  \affil{Department of Mathematics and Statistics, The University of Vermont, Burlington, VT, USA}

\begin{abstract}

   We present deep photometric observations of the open cluster NGC 
2477 using \textit{HST}/WFPC2.  By identifying seven cluster white dwarf candidates, 
we present an analysis of the white dwarf age of this cluster, using both 
the traditional method of fitting isochrones to the white dwarf cooling 
sequence, and by employing a new Bayesian statistical technique that has 
been developed by our group.  This new method performs an objective, 
simultaneous model fit of the cluster and stellar parameters (namely age, 
metallicity, distance, reddening, as well as individual stellar masses, mass 
ratios, and cluster membership) to the photometry.  Based on this analysis, we 
measure a white dwarf age of 1.035 $\pm$ 0.054 $\pm$ 0.087 Gyr (uncertainties 
represent the goodness of model fits and discrepancy among models, 
respectively), in good agreement with the cluster's main sequence turnoff 
age.  This work is part of our ongoing work to calibrate main sequence turnoff 
and white dwarf ages using open clusters, and to improve the precision of 
cluster ages to the $\sim$ 5\% level.

\end{abstract}


\section{Introduction}
   \label{intro}

      Age is one of the most important quantities in astronomy
and is essential to a number of astrophysically interesting problems.  From the
fundamental questions of the formation of the universe to the creation of 
planets, knowing and understanding the ages of celestial objects is 
important.  There are currently two main ways to measure the age of stellar
populations: main-sequence (MS) evolution theory (via cluster isochrones) and 
white dwarf (WD) cooling theory.  These two methods are especially important 
in understanding the ages of the constituents of our own Milky Way.  
Ages determined from the MS turnoffs (MSTOs) of many globular clusters have 
long been used to provide the age of the Galactic halo (e.g., Chaboyer et al. 
1996), while WD cooling ages of field WDs provide the most reliable age of the 
Galactic disk (Winget et al. 1987; Oswalt et al. 1996; Leggett et al. 1998; 
Knox et al. 1999).  


   However, before ages determined by these two techniques can be meaningfully 
compared and the full picture of Galaxy formation understood, they must be 
calibrated to the same absolute scale.  The only way to empirically perform 
this calibration is to measure and compare the ages determined from both 
methods in a single-age, single-metallicity stellar population.  Open clusters 
provide the ideal environment for such a calibration.  In addition to a 
calibration of these two chronometers, measuring the age of a single cluster 
both ways allows us to compare WD theory with MS theory (and vice versa), 
providing an excellent opportunity for the refinement of both.

   Using WDs as the means to measure the age of a stellar population is 
conceptually straightforward.  There are no appreciable internal energy 
sources in a WD (such as nuclear fusion), so it shines because it is hot, and 
as time passes the star cools.  Mestel (1952) first showed that there is a 
relatively simple relationship between the cooling time (i.e., age) of a WD 
and its brightness.  When considered in a simple, conceptual way, this means 
that if we can measure the brightness of the faintest WDs in a stellar 
population, we can measure the population's age.  Although this simplified 
view  of a WD can be helpful in conceptualizing the determination of a WD 
cooling age, in practice, WD cooling models are more complicated than this, 
and measuring an accurate WD age requires other information, including the WDs 
mass.  

   The first studies to apply this technique in open clusters were done by 
Claver (1995) and von Hippel et al. (1995).  Later studies (von Hippel \& 
Gilmore 2000; von Hippel 2001; Claver et al. 2001; DeGennaro et al. 2009; 
Bellini et al. 2010) showed good agreement in WD ages and MS ages for 
clusters up to 4 Gyr.  A summary of these studies and techniques has been 
recently presented by von Hippel (2005).  This technique has also been applied 
to two globular clusters (M4: Hansen et al. 2002; NGC 6397: Hansen et al. 
2007), and a third (47 Tuc, Richer 2010) is underway.  These deep observations 
of the faintest WDs in these globular clusters represents a remarkable 
triumph for observational astronomy.

   While measuring the age of a stellar population with the population's WDs 
seems simple, WD cooling models are complicated.  Substantial work has been 
done to better understand them (e.g., Montgomery et al. 1999, Salaris et al. 
2000), but many puzzles remain, especially among the coolest WDs (e.g., Kilic 
et al. 2006).  This is one of the reasons the studies of WD in star clusters 
is so important.  Studies of WDs in globular clusters allow us to study these 
coolest WDs in a more controlled, uniform population.  Open clusters are also 
important, especially for the calibration of the WD and MSTO ages, as it 
allows us to gradually push up the calibration limit in order to better 
understand the WD cooling physics.  This paper adds to the ongoing effort of 
our and other groups to study WDs in clusters.  Our work complements the 
immense efforts that have gone into the studies of the WDs in the great 
globular clusters.

   An illustration of the need for continuing studies of WDs in open clusters 
and our incomplete understanding of stellar evolution and the formation of WDs 
is the recent case of the old, metal-rich open cluster NGC 6791.  This 
cluster's faint WDs were first observed by Bedin et al. (2005) and initial 
observations showed a peak in the WD luminosity function (LF) that, when fit 
with standard models, indicated an age of 2.4 Gyr, in stark disagreement with 
the MSTO age.  Subsequent studies have sought an explanation for this.  One 
such explanation was put forth by Hansen (2005), who suggested that the bright 
peak was due to a large population of helium-core WDs, formed when stars 
experience excessive mass loss (due to the cluster's high metallicity) along 
the red giant branch (RGB), thereby skipping the helium flash and prematurely 
becoming WDs. Kalirai et al. (2007) used spectroscopic observations to confirm 
the existence of such a population.  Additionally, a second, fainter peak in 
the WDLF of NGC 6791 was discovered with deeper observations (Bedin et al. 
2008b), as Hansen (2005) also predicted.  However, Van Loon et al. (2008), 
using \textit{Spitzer} observations, found no evidence of excessive circumstellar dust 
around the cluster RGB stars, indicating that these stars were not 
experiencing extreme mass loss.  Other groups have also sought to explain the 
strange bimodal WDLF of this cluster, employing scenarios such as high binary 
fraction (Bedin et al. 2008a), and the gravitational settling of $^{22}$Ne in 
the core causing the cooling rate of old metal-rich WDs to be slower than 
normal (Deloye \& Bildsten 2002; Garcia-Berro et al. 2009).

%

   The puzzle of NGC 6791 illustrates the need to continue observations of WDs 
in open clusters.  Our goal is to perform the calibration of MSTO and WD ages 
using open clusters, covering a range of ages and metallicities.  Pushing up 
the limit of calibration gradually will allow us to better understand and 
improve upon any uncertainties in the WD cooling physics, as well as MSTO 
physics.

   This paper presents a study of the open cluster NGC 2477.  We have observed 
it with the \textit{Hubble Space Telescope} (\textit{HST}) to obtain photometry of the cluster 
WDs.  This rich cluster is moderately old (Kassis et al. 1997 measured a MSTO 
age of 1 Gyr), with a distance modulus of 10.61, and differential reddening 
with an average value of $E(B-V)$ = 0.3 (Hartwick et al. 1972).  It is slightly 
metal poor, with [Fe/H] $\sim$ --0.14 (Eigenbrod et al. 2004).  We have listed 
these and several other literature sources for these cluster parameters in 
Table \ref{lit2477}.  This cluster was first studied using \textit{HST} by von Hippel 
et al. (1995), who found a preliminary WD age of 1.0 Gyr.  In a subsequent 
paper, von Hippel et al. (1996) studied the lower mass MS.  In this paper we 
reanalyze these \textit{HST} data and incorporate unpublished Cycle 6 data, using 
updated  \textit{HST} reduction and photometry techniques, new ground-based data for 
the brighter stars, new stellar evolution models, and our new Bayesian 
statistical technique.

   We have organized this paper as follows: Section \ref{obs} outlines our 
observations and data reduction techniques, as well as presenting the deep 
color-magnitude diagram (CMD) for this cluster.  In Section \ref{analysis2477} 
we measure an age for NGC 2477 by fitting WD isochrones to the candidate 
cluster WDs.  We discuss our new Bayesian algorithm in Section \ref{bayes}.  We 
developed this technique to objectively fit models to CMDs and measure cluster 
ages to higher precision than previously possible.  We then apply this 
technique to NGC 2477 and discuss the results in Section \ref{results}.

\section{Observations and Data Reduction}
  \label{obs}

   Two fields of NGC 2477 were observed with the Wide Field Planetary Camera 2 
(WFPC2) aboard \textit{HST}.  Data were taken in two filters, $F$555$W$ and $F$814$W$ 
(equivalent to broadband $V$ and $I$, respectively).  These fields were observed 
on three days over two cycles: 1994 March 18 (Cycle 4), 1996 March 25, and 
1997 March 18 (both Cycle 6).  We have listed coordinate information for the 
two fields in Table \ref{wfpc2_coords_table} and summarized the observations in 
Table \ref{wfpc2_obs_table2477}.

   Basic image calibrations, such as bias correction and flat fielding, were 
performed by the WFPC2 pipeline (Baggett et al. 2002), using the most 
up-to-date calibration files.  We performed the drizzle procedure on the data 
using a detailed cookbook and IRAF\footnote{IRAF is distributed by the 
National Optical Astronomy Observatory, which is operated by the 
Association of Universities for Research in Astronomy, Inc., under cooperative 
agreement with the National Science Foundation.} script provided by STScI in 
the HST Dither Handbook (Koekemoer et al. 2002).

   Sources for photometry were found and morphologically classified using 
SExtractor (Bertin \& Arnouts 1996).  We employed CCDCAP (Mighell \& Rich 
1995) to derive instrumental magnitudes via aperture photometry.  To perform 
aperture photometry, we utilized a stellar aperture of 4 pixels (compare with 
the FWHM of $\sim$ 3.0 pixels for the drizzled images), with a sky annulus of 
radius 10 pixels (0$''$.5), with a width of 3 pixels.  The size of the stellar
aperture was chosen for consistency with the aperture used by Dolphin (2000) 
when calculating charge transfer efficiency (CTE) corrections.  We used these 
corrections from Dolphin (2000) with the most up-to-date corrections 
available on his Web site\footnote{\url{http://purcell.as.arizona.edu/wfpc2\_calib/}}.
The corrections we used here were taken from the 2008 October update of his 
Web site.

   The zero points (ZPs) and color terms we used were those determined by 
Holtzmann et al. (1995) and updated by Dolphin (2000).  We used the most 
up-to-date values (available from Dolphin's Web site) to transform raw, 
CTE-corrected WFPC2 magnitudes to standard $V$ and $I$ magnitudes.  We note that 
the cluster observations were taken with the gain setting of seven, so the 
appropriate ZPs were used.  We also applied an aperture correction to correct 
the smaller aperture measurements of the observations to the appropriate 
size of the standards.

\begin{figure}[h!]
   \epsscale{1.00}
        \plotone{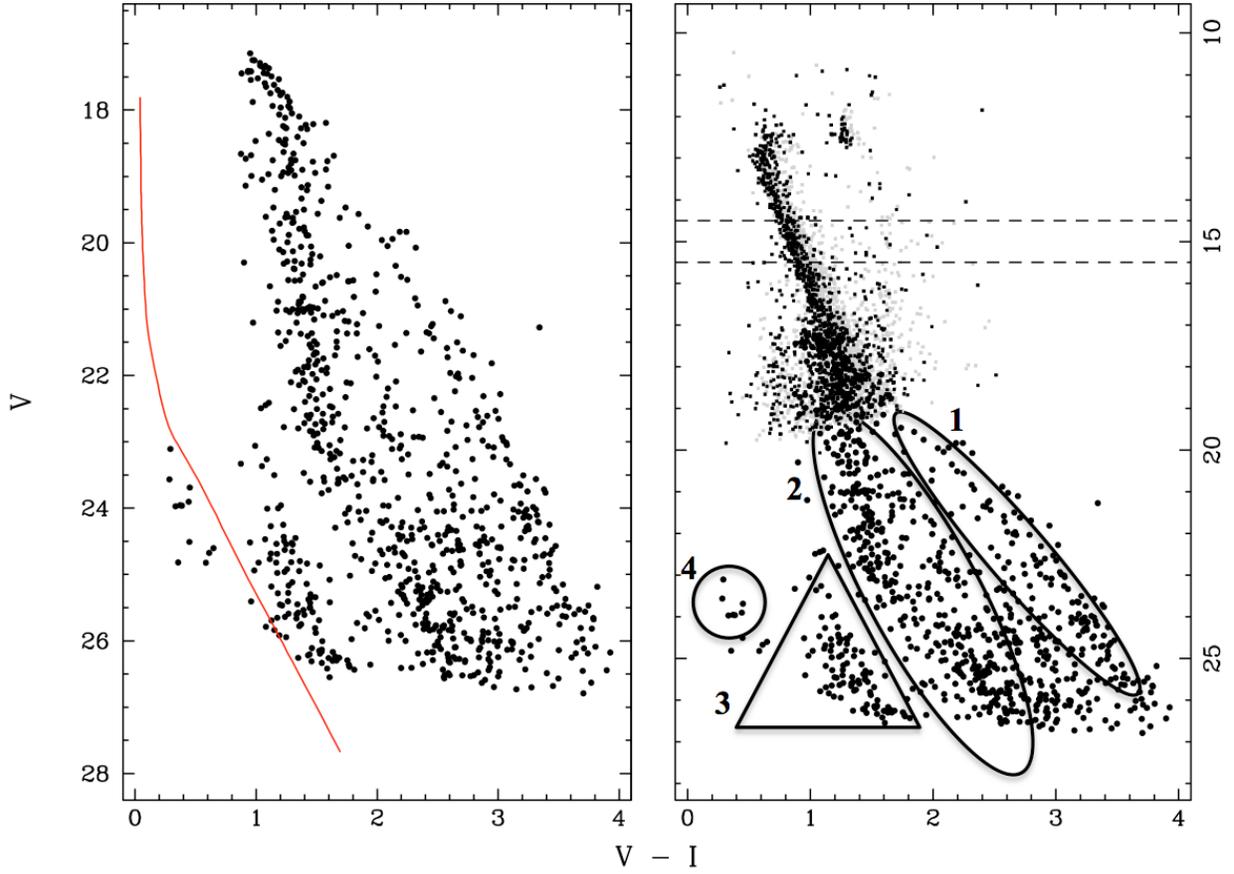}
        \caption{Left: CMD of NGC 2477 taken with HST/WFPC2.  Only objects 
		with magnitude errors less than 0.1 have been plotted.  
                Overplotted is the cooling track of a 0.6$M_{\odot}$ WD 
                (Wood 1992). Right: Ground-based photometry (small black and 
		gray points) of NGC 2477 combined with HST data (large black 
		points).  The numbered regions indicate 
		(1) the cluster MS; (2) a substantial stellar background 
		population; (3) other field stars, image defects, and 
		background galaxies; and (4) the cluster WD sequence.  The 
		dashed lines in this panel indicate 
		the portion of the MS used when running MCMC (see Section 
		\ref{bayes_prep}).  That is, when running MCMC, \textit{only} 
		the portion of the MS between the dashed lines as well as the 
		cluster WDs candidates are used.  The rest of the data are 
		discarded.}
   \label{hst_cmd2477}
\end{figure}

   In the left panel of Figure \ref{hst_cmd2477}, we present the CMD of all the 
objects in our WFPC2/\textit{HST} fields, with only objects with magnitude ($V$ and $I$) 
errors less than 0.1 plotted.  Overplotted in this figure is the cooling track 
of a 0.6$M_{\odot}$ WD (Wood 1992), assuming a true distance modulus (i.e., 
$(m-M)_{0}$) of 10.74 and reddening ($E(B-V)$) of 0.23, Salaris et al. 2004).  
We note that the average WD mass of this cluster is likely higher than 
0.6$M_{\odot}$; however, since this is roughly the average of field WDs 
(e.g., Kepler et al. 2007), it serves as a good first approximation in helping 
us locate the position of the WD sequence.

   The right panel of Figure \ref{hst_cmd2477} combines the \textit{HST} data (large 
black points) with a set of new photometric observations of the upper MS of 
NGC 2477 (small black and gray points).  (The black points are those that lie 
within 300 arcsec of the cluster center, and are therefore more likely to 
be cluster members, while the gray points lie outside the cluster core.)  The 
details of these data will be forthcoming in a later paper (E. J. Jeffery et al., in 
preparation).  Briefly, we observed NGC 2477 with the SMARTS 1-m telescope and 
Y4KCam CCD camera\footnote{More information about the Small and Moderate 
Aperture Research Telescope System (SMARTS) can be found at \url{http://www.astro.yale.edu/smarts/}} 
at Cerro Tololo Inter-American Observatory.  The CCD provides a 
20$\times$20 arcmin field of view with a scale of 0.298 arcsec 
per pixel.  Observations were taken with standard $BVI$ filters and achieve a 
signal-to-noise ratio of 5-10 at $V \sim$ 19.5.  In the current analysis, the 
data in $V$ and $I$ will be used in combination with the \textit{HST} photometry of the 
cluster WDs.

   Because the CMD of this cluster is somewhat complicated, we have classified 
it into four main parts, as diagramed in the right panel of Figure 
\ref{hst_cmd2477}.  Region 1 is the cluster MS.  Region 2 is a significant 
background population that has merited some discussion in the literature 
(e.g., Momany et al. 2001); some argue that it is part of the Canis Major 
overdensity system (Bellazzini et al. 2004), while others argue that it is not 
part of the CMa system (Carroro et al. 2005).  Nonetheless, it is well accepted 
that this background population exists and that it is not associated with NGC 
2477.  Region 3 includes other field stars, image defects, and background 
galaxies.  And finally, Region 4 is the cluster WD cooling sequence.

   Also included in the right panel of Figure \ref{hst_cmd2477} are two 
horizontal dashed lines on the upper MS.  This indicates the region of the MS 
used in our Bayesian analysis and will discussed further in Section 
\ref{bayes_prep}.

    Our observations were taken at multiple epochs (see Table 
\ref{wfpc2_obs_table2477}), so we investigated the feasibility of measuring 
proper motions (PMs) in order to clean the CMD.  PMs would be extremely useful 
for separating the cluster members from the field (especially with such a 
large  background field population).  The observations for Field 2 were taken 
on the same day, making measurements of PMs impossible.  The observations of 
Field 1 were taken just two years apart.  Dias et al. (2006) measured a PM for 
this cluster as differing from the field by less than 2 mas year$^{-1}$.  Given this, 
we would expect $\sim$ 3.5 mas of movement over the two-year baseline, translating 
to 0.07 WFPC2 drizzled pixels.  

   Because we are reanalyzing data taken during Cycle 4 and 6, i.e., taken 
before we properly understood subpixel offsets and drizzling procedures, the 
resulting drizzled point spread functions (PSFs) are not as well constructed when compared to current 
standards, simply because of the way the data were taken.  The result is that 
we cannot calculate the centroids to the precision required to measure the PM 
of this cluster.  Fortunately, we can use morphological information to reject 
many of the background galaxies (see Section 3), allowing us to proceed in our 
analysis without PM information.


\section{Fitting White Dwarf Isochrones}
   \label{analysis2477}

\begin{figure}
   \epsscale{1.0}
        \plotfiddle{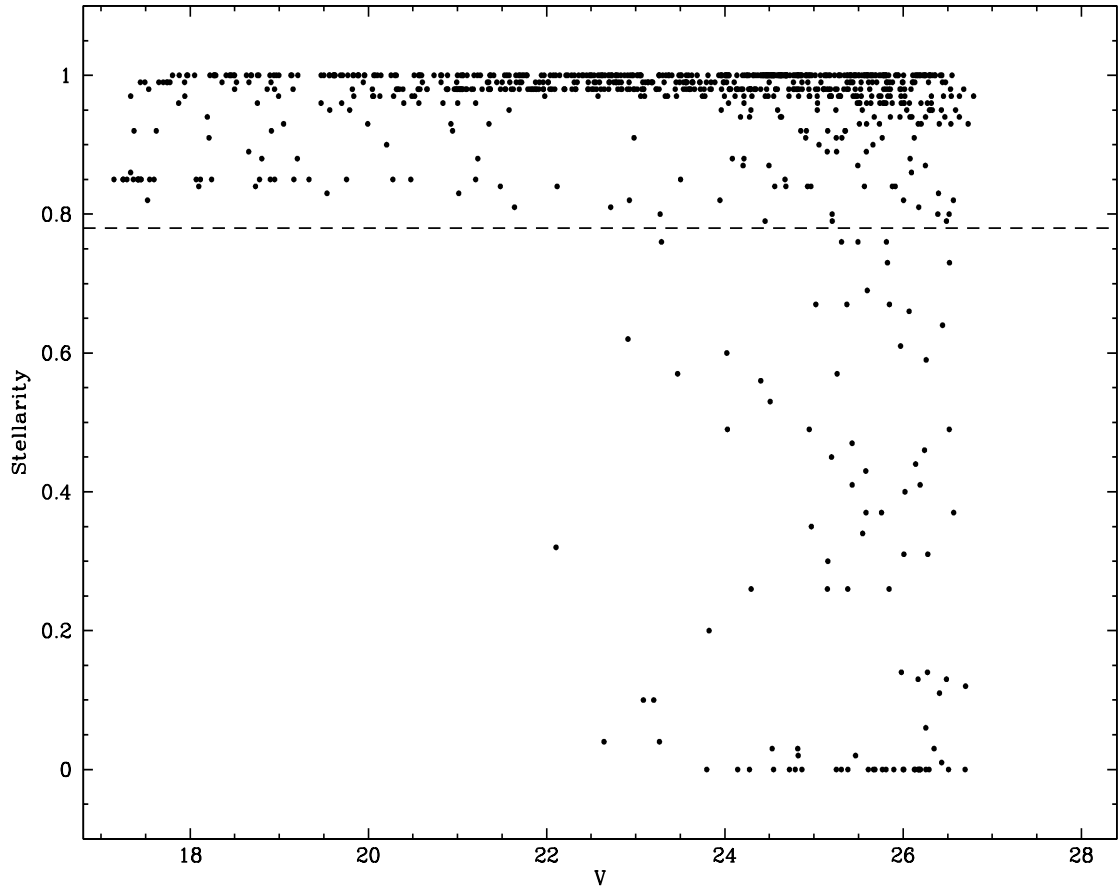}{0.0in}{-90}{376}{466}{0}{25}
      \caption{Stellarity index vs. V magnitude.  Stellarity index ranges
                from 0 (galaxies) to 1 (stars).  Objects with stellarity $\le$
                0.78 (indicated by the dashed line) were automatically 
		rejected from the analysis.}
      \label{stel}
 \end{figure}

   Despite contamination in the region of the cluster WDs being minimal (see 
Figure \ref{hst_cmd2477}), star-galaxy separation was still an important point 
of concern.  Because of small shifts among the images input to the drizzle 
algorithm, SExtractor had a difficult time distinguishing between stars and 
galaxies on the drizzled images.  To mitigate this problem, we ran SExtractor 
on images that were a simple combination of individual images (using the IRAF 
IMCOMBINE task, rather than drizzle), and matched the object classification 
with the master (drizzled) photometry values.  We present a plot of stellarity 
versus magnitude in Figure \ref{stel}.  The so-called ``stellarity index'' ranges 
from 0 (galaxies) to 1 (stars).  All sources with a stellarity index less than 
0.78 were automatically rejected.  This threshold was chosen based on results 
from von Hippel \& Gilmore (2000), although the final WD candidates all had 
stellarity indices greater than 0.90 (i.e., they are definitely stars).  The 
stellarity values provided the first cut toward isolating a clean stellar 
sample of cluster WDs candidates.

\begin{figure}
   \epsscale{1.0}
        \plotfiddle{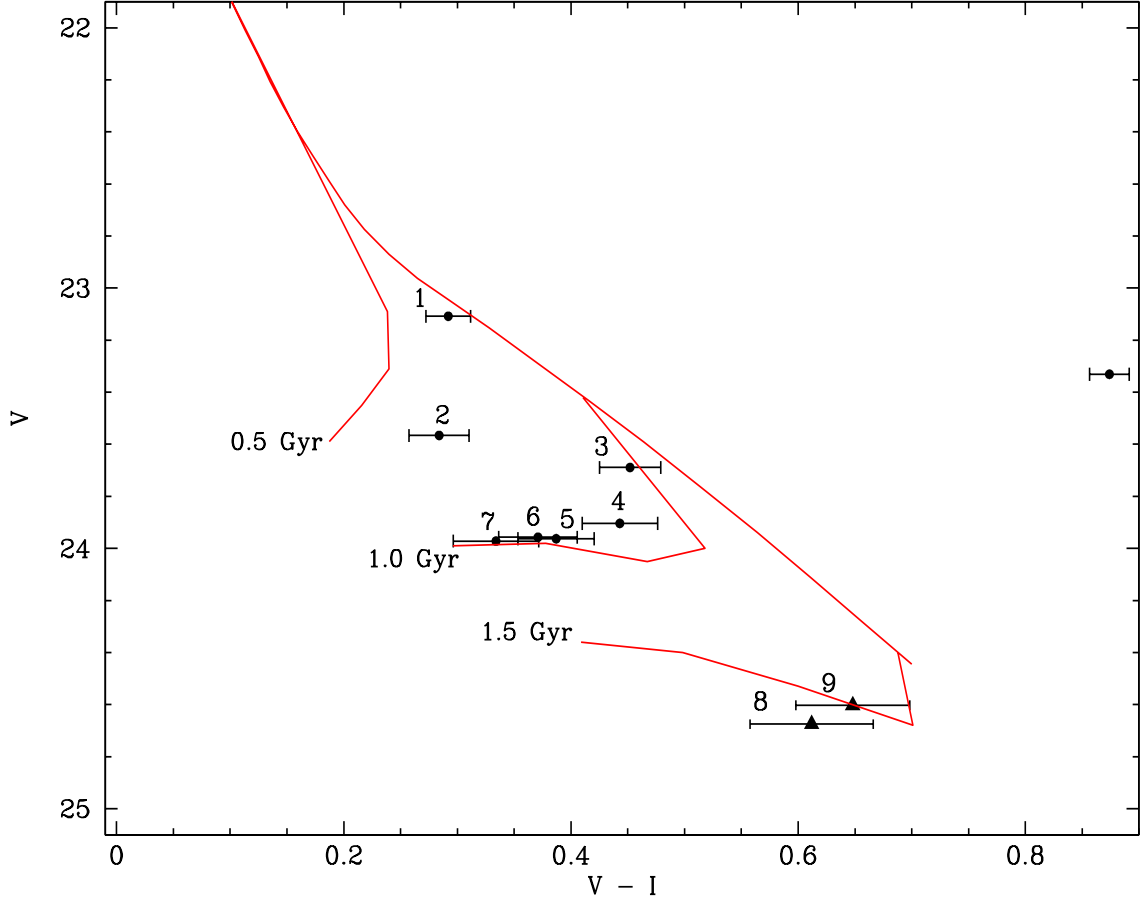}{0.0in}{-90}{376}{466}{0}{25}
        \caption{CMD of NGC 2477, zoomed on the WD region.  Only objects 
		passing the first stellarity cut (Figure \ref{stel}) are 
		plotted.  Each object was then visually inspected on the 
		original images.  Objects 1-7 are confirmed to be stellar and 
		we assume them to be the cluster WDs; objects 8 and 9, 
		plotted as triangles, are found to be image defects (see 
		Figure \ref{starcont}) and will be discarded from further 
		analysis.  We have overplotted WD isochrones (from Wood 1992) 
		for 0.5, 1.0, and 1.5 Gyr. The best fit to the WDs (i.e., 
		objects 1-7) in this cluster is 1.0 Gyr, in good agreement 
		with the MSTO age.}  
\label{ngc2477_hst_wds}
\end{figure}

\begin{figure}
   \epsscale{0.95}
        \plotone{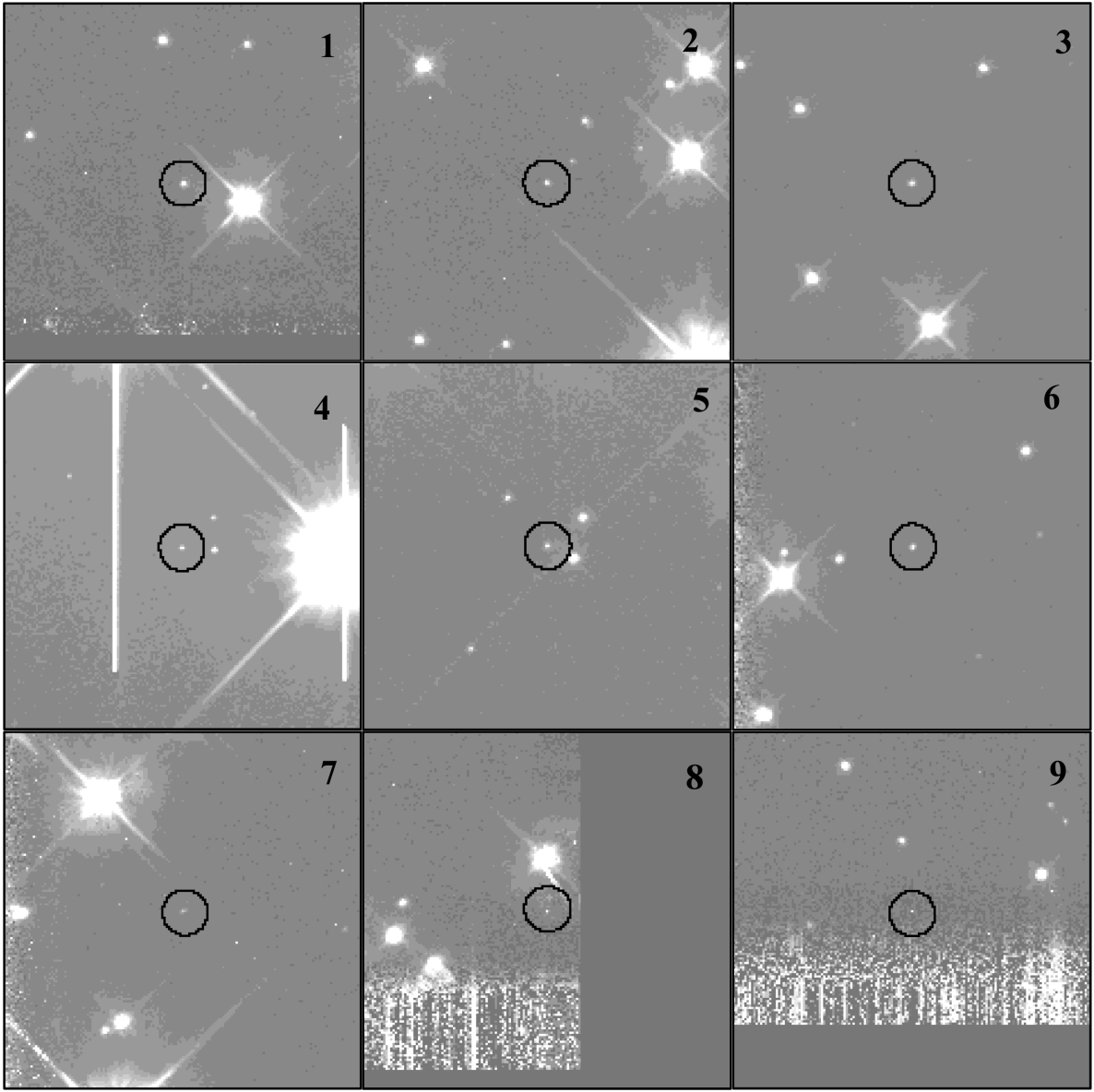}
        \caption{Image screenshots of the nine blue objects in the WD region 
		of the CMD with high stellarity (objects 1 through 9, see 
		Figure \ref{ngc2477_hst_wds}).  (We note that we are 
		displaying the imcombined, not drizzled, images.  See text for 
		further discussion.)  Visual inspection confirms that objects 
		1 - 7 are indeed stellar, while 8 and 9 are they are likely 
		image defects, found in the noisy vignetted edge region of the 
		CCD.  Because of this, these two objects will be completely 
		discarded for further analysis.} 
   \label{starcont}
\end{figure}

   After the stellarity cut, nine possible WD candidates remained.  We plot 
these in Figure \ref{ngc2477_hst_wds}, a zoomed region of the CMD around the 
region of the WDs.  We visually inspected each object on the original images 
to confirm the stellar nature of each one and to exclude any image defects 
(hot pixels, diffraction spikes, etc) or other non-stellar objects.  In Figure 
\ref{starcont}, we display the portion of the image used for visual inspection 
with each of these nine possible WD candidates indicated.  This examination 
confirms that objects 1--7 are stellar, while 8 and 9 are likely image defects, 
found in the noisy vignetted edge region of the CCD.  Because of this, these 
two objects were discarded from further analysis.  In our first analysis, using 
standard WD isochrone fitting, we assume that the stellar objects in the WD 
region are primarily cluster WDs.  We relax this assumption in our full 
Bayesian analysis, as mentioned below.

   As an important consistency check, we calculated how many WDs we would 
expect to find, based on the number of stars we observe on the MS.  To do this 
calculation, we first count the number of MS stars in a particular magnitude 
range.  (We chose bright MS stars, so as to avoid the complications of 
incompleteness.)  We then simulate a cluster with the previously published age 
and metallicity of NGC 2477, incorporating a Miller \& Scalo initial mass 
function (IMF), MS evolution timescale models of Dotter et al. (2008), the 
initial--final mass relation of Weidemann (2000), WD cooling timescales of Wood 
(1992), and WD atmospheres color from Bergeron et al. (1995).  Because this 
process can be somewhat stochastic, we produced 1000 clusters in this manner 
and selected only those clusters with a number of MS stars in the same 
magnitude bin equal to that determined for NGC 2477, totaling 165 simulated 
clusters.  From these 165 clusters, we calculated the average and standard 
deviation of the number of WDs produced in each instance to be  6.5 $\pm$ 
2.4.  This is a good estimate of the number of WDs we expect to find.  Our 
finding of seven WD candidates falls within the expected number.   

   We also expect the number of field WDs or unresolved quasars in this region 
to be small.  To estimate the total number of background quasars, we first take 
the observed number denisty of these objects from recent Sloan Digital Sky 
Survey observations (e.g., Richards et al. 2009) and use a quasar LF (e.g., 
Richards et al. 2006) to extend this number density from their limiting 
magnitude to that of our cluster WD terminus.  By then scaling this number 
density to the field of view of our observations, we expect to find less than 
one quasar in our field.  We therefore do not expect background quasars to be a 
problem and proceed with our analysis.

   In many cases when observing the faintest members of a cluster, 
observational incompleteness becomes an important issue.  von Hippel et al. 
(1996) calculated the incompleteness of Field 1 to be $\sim$98\% at the 
level of the coolest WD candidates (see their Figure 1).  Field 2 was taken 
with similar exposure times, so we can assume the completeness to be similar.  
Because of this high completeness, the data do not need to be corrected for 
observational incompleteness.

   We fit isochrones to the candidate WDs to estimate the WD age for NGC 
2477.  In Figure \ref{ngc2477_hst_wds}, we overplot several Wood (1992) 
isochrones for ages of 0.5, 1.0, and 1.5 Gyr, as labeled.  Objects 1 through 7 
are our WD candidates.  We have listed the photometry and coordinate positions 
for these stars in Table \ref{ngc2477_wds_table}.  This demonstrates a best 
fit to the terminus of the WD cooling sequence with the 1.0 Gyr isochrone.  
This is in good agreement with the MSTO age (Kassis et al. 1997).  We will 
discuss error analysis in a later section.


\section{A Bayesian Approach to Measuring Cluster WD Ages}
  \label{bayes}

   The method used in the previous section for determining cluster ages (that 
is, fitting isochrones to the WD cooling sequence) has served us well for 
years.  However, we desire a more objective approach to fitting models to 
the data, as well as to increase the precision of the fit.  Our goal is to 
improve the precision of both MSTO and WD ages to 5\%.  Because no qualitative 
gains are likely to be made in the near future in the precision of cluster 
photometry, stellar abundances, or cluster distances, our best hope of 
achieving this high age precision is from improved models and improved 
statistical fitting procedures.

   In this section, we will discuss the application of a powerful new Bayesian 
technique we have developed to determine cluster ages with higher precision.  
We give a brief overview of the technique and then apply it to the data set 
discussed in Section \ref{obs} to measure the WD age of NGC 2477 from the 
cluster WDs.

\subsection{Overview of the Technique}
   \label{overview}

   Our Bayesian technique derives posterior distributions for various cluster 
and stellar parameters by utilizing Bayesian analysis methods.  These 
parameters include age, distance modulus, metallicity, and line-of-site 
reddening, as well as individual stellar masses, mass ratios, and cluster 
memberships probabilities.  For a thorough and in depth discussion of the 
technique and its first applications, we refer the reader to von Hippel et al. 
(2006), Jeffery et al. (2007), DeGennaro et al. (2009) and van Dyk et al. 
(2009).  However, for the convenience of the reader, we provide a short 
overview of the technique here.

   Our Bayesian algorithm aims to fit a cluster evolution model using clear 
statistical principles.  In our current analysis, the cluster evolution model 
is as follows.  It incorporates a Miller \& Scalo (1979) IMF, one of three 
available MS evolution timescale models (namely, the 
models of Dotter et al. 2008 (DSED), Yi et al. 2001 (YY), and Girardi et al. 
2000), the initial (MS) mass--final (WD) mass relation of Weidemann (2000), WD 
cooling timescales of Wood (1992), and WD atmospheres colors from Bergeron et 
al. (1995).  In our current analysis, we will compare and discuss the results 
of using each of the three MS evolution timescale models, while the other model 
ingredients are held fixed.

   Bayes' theorem is at the heart of Bayesian statistics.  It states that 
the posterior probability distribution of the parameters of our model (e.g., 
cluster age) is proportional to the product of the prior probability 
distribution and the likelihood function.  The prior probability distribution 
incorporates any information from outside the data, including the 
cluster-wide parameters age, metallicity, distance, and reddening, and 
individual stellar parameters such as cluster membership probability (if 
available from either PMs or radial velocity measurements), mass 
determinations, and the mass ratio of any unresolved companions.  The 
likelihood function compares the predicted photometry for each star (via our 
cluster evolution model) with the observed photometry assuming known errors in 
the latter.

   Because of the high dimensionality and complex nature of these 
distributions, the equations cannot be manipulated analytically.  We use a 
Markov chain Monte Carlo (MCMC) technique to sample the posterior 
distributions of the different cluster-wide parameters, as well as individual 
stellar quantities such as the mass of each star, mass ratio of any unresolved 
binary companions, and membership probability.  The convergent MCMC chain 
provides a (correlated) sample from the posterior distributions of the 
corresponding quantities, and can be used to compute means and intervals as 
parameter estimates and error bars.



   We wish to note that the precision of the parameters discussed in the 
following sections is internal precision, rather than external accuracy.  Our 
technique objectively determines the best fit of the model to the data, and 
the goodness of that fit; it cannot assess the physical accuracy of the model 
itself.  For example, a change in the initial-final mass relation could change 
our results systematically, but the internal precision would still be high.  A 
better understanding of such external modeling issues comes as we intercompare 
results from multiple model sets, as we begin to do here.

\subsection{Input Data}
  \label{bayes_prep}

   For the application of the Bayesian technique\footnote{For ease, from this 
point forward we will refer to the Bayesian algorithm simply as ``MCMC."}, we 
use the photometry of the seven WD candidates (see Table 
\ref{ngc2477_wds_table}) and ground based photometry of the cluster MS (taken 
from E.J. Jeffery et al. in preparation, as discussed in Section \ref{obs}).  Only a 
small portion of the MS is used; for this cluster we used the MS between 
$V$ = 14.5 to 15.5.  Assuming a reddened distance modulus of 11.45 for this 
cluster, which translates to an absolute magnitude of 3.05 -- 4.05\footnote{The 
exact limits on the MS are relatively unimportant, as long as the criteria of 
the upper and lower limits, discussed in this section, are met.}, or a mass 
limit of 1.16 -- 1.40$M_{\odot}$, we have indicated these limits with the 
dashed horizontal lines in Figure \ref{hst_cmd2477}.  (We note that everything 
outside the dashed horizontal lines, except the seven WD candidates, was 
discarded from the Bayesian analysis.)  The upper MS magnitude cutoff was 
imposed to remove the entire turn off region, in order to derive age 
information from the WDs alone.

   A lower MS limit was imposed to control the amount of MS to be included in 
each run.  The MS fit provides the primary constraint on cluster metallicity, 
distance, and reddening.  While models tend to fit the upper MS well, most do 
a poor job at fitting the lower MS.  Because the MCMC fit is predicated upon 
the cluster evolution model, the poor fit of the model to the data can result 
in poor fits of certain cluster parameters.  (This specific issue was explored 
extensively by DeGennaro et al. 2009.)  To bypass this problem, we only use 
the upper MS (below the MSTO) where the models better fit the observed shape 
of the cluster MS.  For this reason, the MS observed from the deep \textit{HST}  
photometry was not used.  Rather, MCMC was given the WD photometry from \textit{HST}  
observations and a MS from the ground based data from E.J. Jeffery et al. (in preparation).

\subsection{Running MCMC}
   \label{priors}


   For each of the three MS stellar model sets, we ran a total of twelve 
separate MCMC chains.  Each chain was set to sample for $2\times10^{6}$ 
iterations, reading out every 100th iteration, until 20,000 values of the 
posterior distribution have been saved.  Prior to each run was a burn-in 
period.  During this burn-in period, which totaled 130,000 samples per chain, 
the MCMC chain was allowed to stabilize, and various correlations were calculated 
(see DeGennaro et al. 2009 for more details).  This allows for more efficient 
sampling during the MCMC run.  We note that all statistics were calculated 
after the burn-in.

   We initiated each of the chains at the prior means of metallicity, distance 
and reddening (see Table \ref{priors2477}).  We used a flat prior on 
log(age).  To check for sensitivity to the starting value of this parameter of 
primary interest, as well as to look for different modes in the posterior 
distribution (i.e., competing best/good fits), we started chains at three 
different starting values for age, namely log(age)=8.9, 9.0, and 9.1.  
Starting values for the masses were compiled by creating an isochrone at the 
starting value of the cluster parameters and computing the mass for each star 
that results in the best match between its observed colors and magnitudes and 
the created isochrone.  Because no additional information was available (from 
radial velocity data or PMs), we set the starting values for the 
mass ratios to 0.1 and cluster membership priors to 0.5.  All of this results 
in three chains for each MS model, for a total of nine chains.  As we describe 
below, this entire scenario was replicated four times using different random 
seeds, or twelve chains for each MS model, for a grand total of 36 MCMC chains.

\begin{figure}
   \epsscale{0.85}
        \plotone{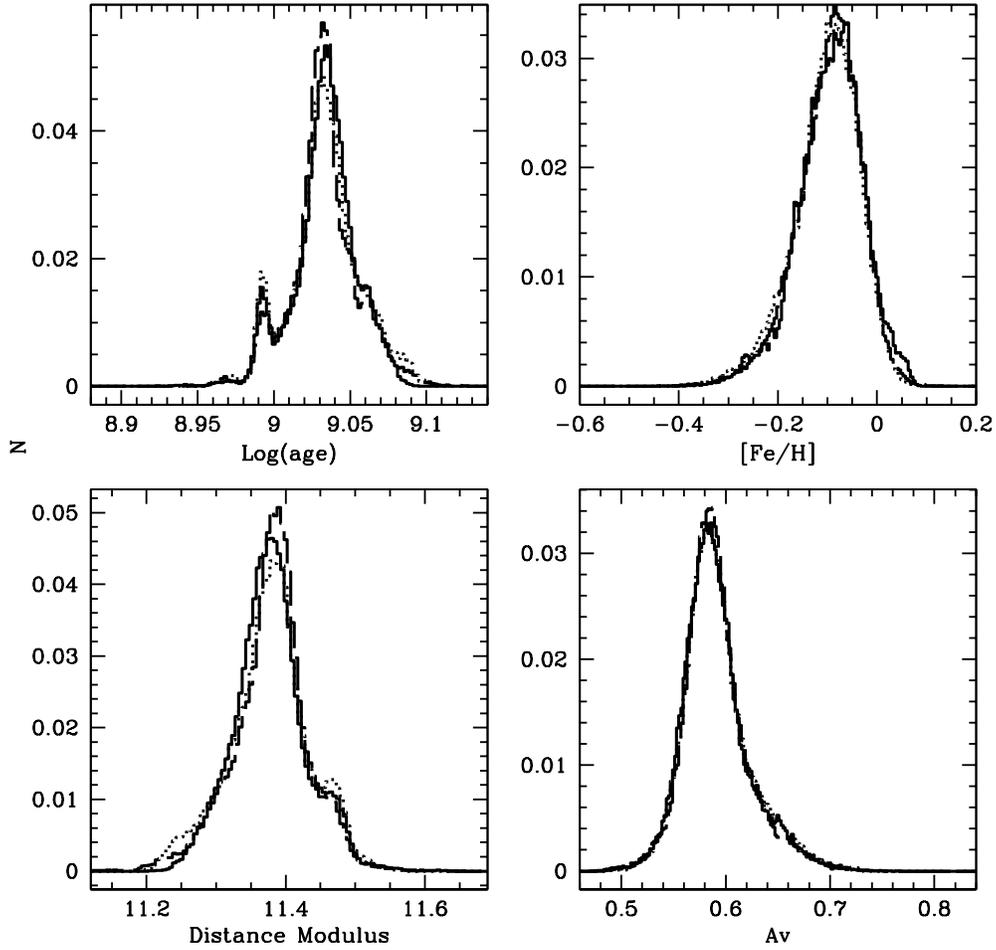}
        \caption{Comparison of posterior distributions of cluster parameters 
		given different start values for log(age), namely log(age) = 
		8.9, 9.0, and 9.1, represented by different line styles: solid 
		(8.9), dashed (9.0), and dotted (9.1).  We note how 
		consistently MCMC found the posterior distribution, regardless 
		of starting value.  This demonstrates the robustness of the 
		technique to the age starting value, and the plausibility of 
		no additional modes in this area of parameter space.}
   \label{compstart}
\end{figure}

   We compared the results for the three starting log(age) values for each 
model, and in each case, MCMC consistently found the same location of the 
posterior distribution.  This demonstrates both the robustness of the 
algorithm on determining the posterior distribution, as well as indicate the 
plausibility that no other modes in the posterior distribution exist in this 
region of parameter space.  We plot this result in Figure \ref{compstart} for 
the DSED models, but note similar results for both the YY and Girardi models.  
In this plot, we show the sampling histories of these runs as 
histograms.  Each panel is a different cluster parameter (as labeled).  The 
three line styles represent the three start values of log(age): solid (8.9), 
dashed (9.0), and dotted (9.1).  Clearly, MCMC finds the same posterior 
distribution, regardless of the start value. 

\begin{figure}
   \epsscale{0.85}
        \plotone{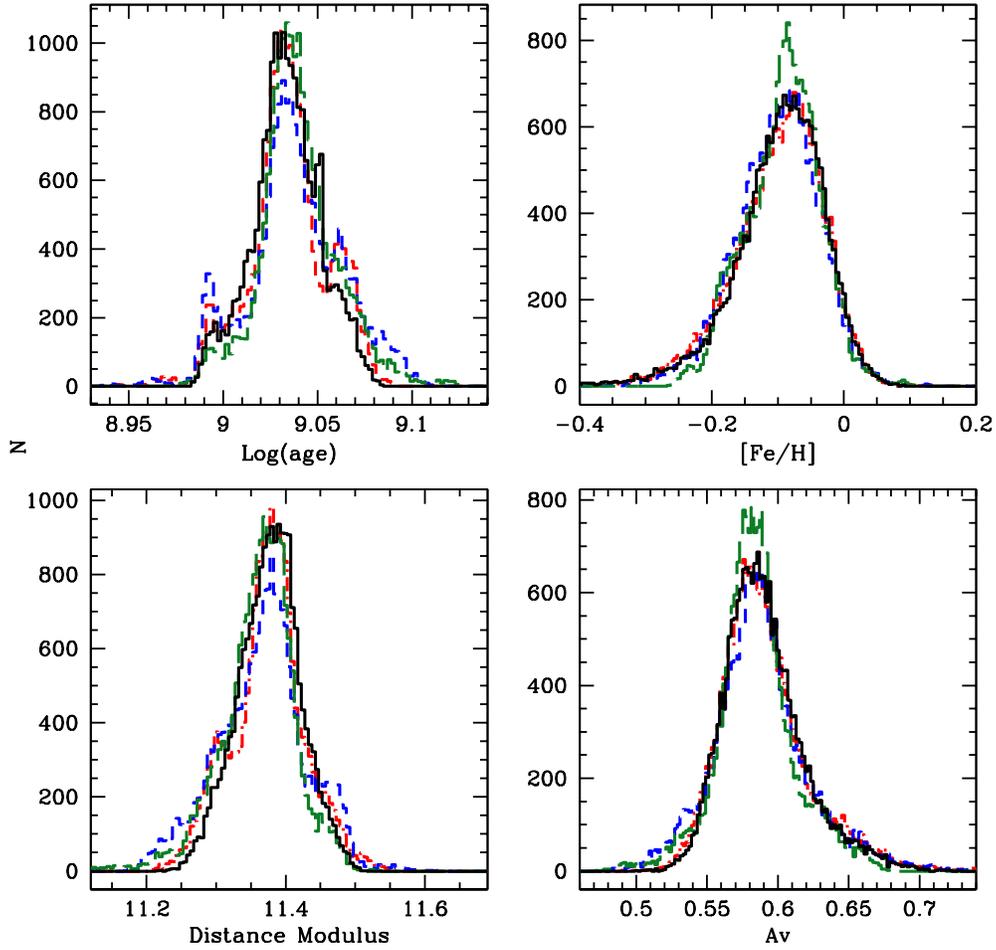}
        \caption{Histograms of the sampling histories for the cluster 
		parameters of NGC 2477 using different initial random seeds 
		for each run.  The results plotted here used the DSED MS 
		models with a log(age) starting value of 9.0 (see Section 
		\ref{priors}).  Different runs using different random seeds 
		are represented by different colors and line styles.  From 
		this it can be seen how consistently MCMC found the location 
		of the posterior distribution for sampling, regardless of the 
		initial random seed.}
   \label{overplot_sampling_2477_hist}
\end{figure}

   For each combination of stellar evolution model and starting value for 
log(age), we set MCMC to run four times, using four different initial seeds.  
Each time, the MCMC chain stabilized and sampled well.  We demonstrate this in 
Figure \ref{overplot_sampling_2477_hist}.  This represents multiple runs 
starting at log(age) = 9.0, using the DSED models, although results were 
similar for both the YY and Girardi models, as well as for other log(age) 
start values. This demonstrates that the posterior distribution is located and 
well sampled, regardless of the initial random seed.

\section{Results}
   \label{results}

   Once we had the full posterior distributions, we were able to calculate 
statistics.  We emphasize that the best summary of our analysis is the complete 
posterior distribution.  Because of its high dimensionality (four cluster 
parameters plus three parameters for each star), we focus on simpler and more 
familiar summaries that are easier to compute.  In particular, we report the 
average and the standard deviation of the sample MCMC chains combined from 
individual runs with each of the three starting values for log(age) and each 
of the four random seeds.  Again, we emphasize that the precision reported 
here is internal precision.

\begin{figure}
   \epsscale{0.85}
        \plotone{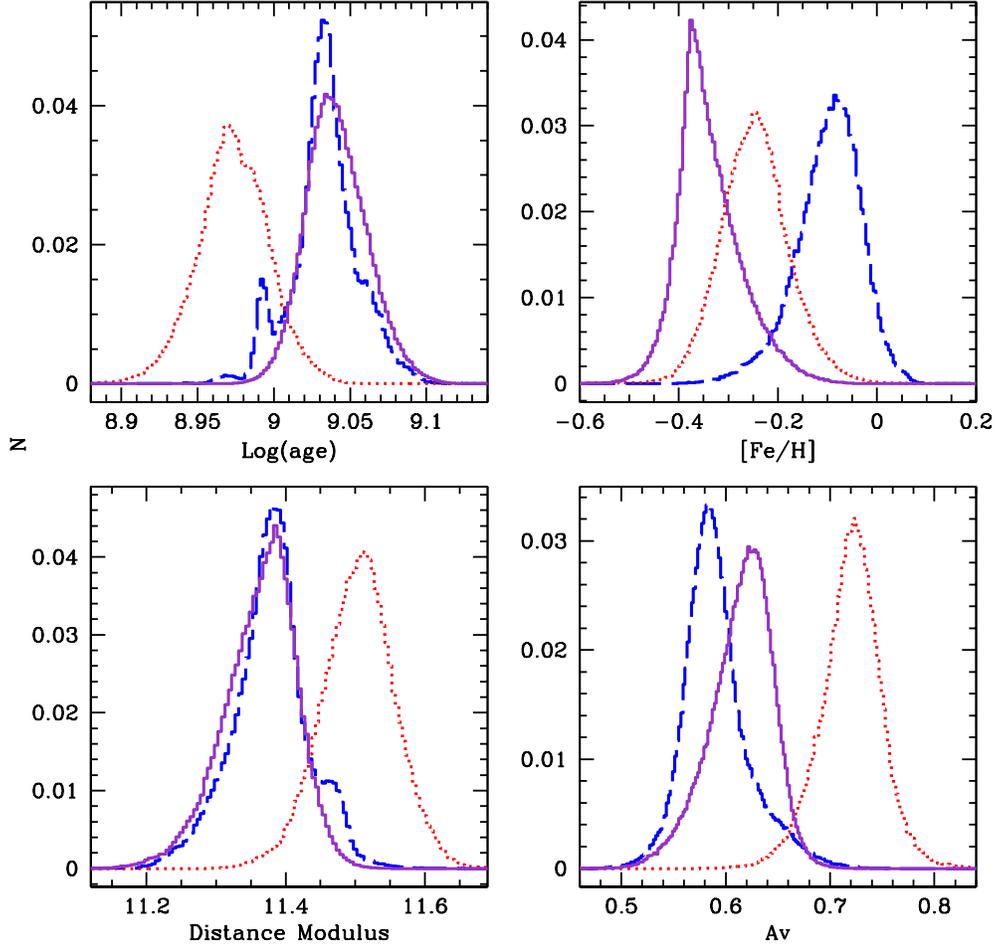}
        \caption{Comparison of the combined posterior distributions of cluster 
		parameters given the different MS evolution models.  Each 
		posterior distribution is the combination of twelve individual 
		MCMC runs (i.e., three different log(age) starting values, 
		each run with four different random seeds).  The different 
		models are represented by different line styles: solid purple 
		(Girardi et al. 2000), dotted red (Yi et al. 2001), and dashed 
		blue (Dotter et al. 2008).  For age, DSED and Girardi models 
		are in the best agreement, while the YY models give a slightly 
		lower age, although the distributions are all overlapping.  We 
		have listed the average and standard deviation for each of 
		these distributions in Table \ref{bayes2477wd_results}.}
   \label{comp_models}
\end{figure}

   We compare the values determined by each of the MS evolution model sets we 
used.  In Figure \ref{comp_models}, we overplot the full posterior 
distributions of each model set, a combination of the twelve individual runs 
for each model set, as discussed in Section \ref{priors}.  Colors and line 
styles consistent with DeGennaro et al. (2009), namely solid purple (Girardi 
et al. 2000), dotted red (Yi et al. 2001), and dashed blue (Dotter et al. 
2008).  The average and standard deviation for each model set is listed in 
Table \ref{bayes2477wd_results}.  The DSED and Girardi values agree well on 
the age, while the YY models give a slightly younger age, although the 
distributions still overlap.

   As we explained in Section \ref{overview}, MCMC compares the observed 
photometry to the predicted photometry of each star given the set of stellar 
cluster parameters for a given step in the MCMC chains.  The predicted 
magnitudes of each star can then be averaged over the course of the chain and 
compared to the data.  This demonstrates how well MCMC models the photometry 
and indicates the reliability of the fit.  

\begin{figure}
   \epsscale{0.90}
        \plotfiddle{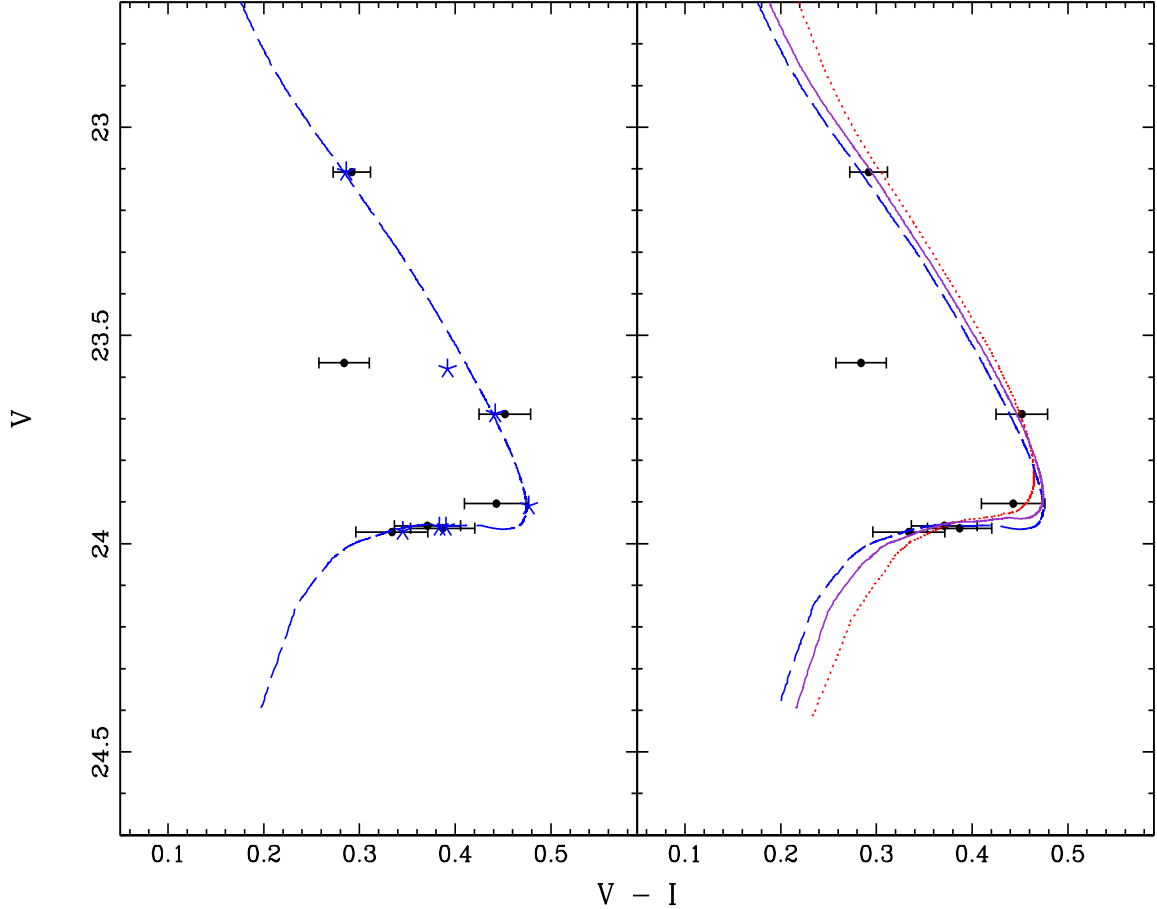}{0.0in}{-90}{376}{466}{0}{25}
        \caption{The CMD of the WD region of NGC 2477.  In both panels, the 
		solid black points are the data.  On the left, the blue 
		stars represent the average of the photometry values produced 
		by MCMC (using the DSED models).  A WD 
		isochrone simulated with the DSED values in Table 
		\ref{bayes2477wd_results} is overplotted to demonstrate the 
		quality of the fit to the data.  (We also note that for WD2 in 
		this panel, the final cluster membership probability from each 
		MCMC run was typically low, often less than 10\%, which 
		explains for the offset predicted photometry value.)  
		The right panel of this plot shows multiple WD isochrones 
		simulated with the appropriate values for that model, listed 
		in Table \ref{bayes2477wd_results}, for easy comparison of the 
		models in color-magnitude space.  Color and line styles are 
		consistent with those used in Figure \ref{comp_models}.} 
   \label{bayes2477_wd_outcmd}
\end{figure}

   In both panels of Figure \ref{bayes2477_wd_outcmd}, we display the CMD of 
just the WD region of NGC 2477, with the photometry of the WD candidates given 
by the black points and error bars.  In the left panel, we plot the average 
values of the photometry as predicted by MCMC for the DSED model set, 
represented by the blue stars.  We note that for WD2, the final cluster 
membership probability from each MCMC run was typically low, often less than 
10\%, explaining for the somewhat skewed predicted photometry.  Also 
overplotted is a WD isochrone, simulated with the best fit cluster parameters 
(by MCMC using the DSED models) listed in Table \ref{bayes2477wd_results}.  
This figure demonstrates that the photometry data were fit very well and that 
the combination of parameters found (especially age) by MCMC did an excellent 
job at fitting the data.  The right panel of this plot shows multiple WD 
isochrones simulated with the values in Table \ref{bayes2477wd_results} for 
all models, for easy comparison of the models in color--magnitude space.  

   Our final value for the WD age of this cluster comes by taking a weighted 
average of the ages determined by all three model sets, calculated in the 
standard way, with weights being inversely proportional to the MS model 
specific variances.  We then calculated two error bars.  The first represents 
a quality of the model fit to the data (i.e., internal or within-model error) 
and is found by taking the square root of the average of the variances for 
each MS model (i.e., the square of the standard deviations, as listed in Table 
\ref{bayes2477wd_results}).  The second represents the spread among the three 
different models we used, found by computing the standard deviation of the 
three model-specific fitted values (i.e., average of the MCMC chains, see 
Table \ref{bayes2477wd_results}).  It is a preliminary measurement of the 
theoretical uncertainty in the fits, to the degree that these three models 
span the range in uncertainty in the input physics.  (Although they do not 
capture all of that uncertainty, it is a start in the process of quantifying 
it.)

   With this in mind, our final value for the WD age of NGC 2477 is 1.035 
$\pm$ 0.054 $\pm$ 0.087 Gyr.

\section{Conclusions}
  \label{conclusion}

   In this paper, we have presented deep \textit{HST} observations of NGC 2477.  We have 
estimated a WD age of this cluster using traditional techniques of fitting WD 
isochrones by eye, resulting in a WD age of approximately 1.0 Gyr.  In order 
to achieve higher precision in clusters ages, our group has developed a new 
algorithm that utilizes Bayesian statistics.  By employing an MCMC  
technique, we are able to sample from the posterior distributions 
of cluster parameters, specifically age, performing a simultaneous best fit of 
the free parameters to the photometry.  WD ages determined from this method 
have much higher precision than by simply fitting isochrones by eye.  Using 
the Bayesian algorithm and utilizing different MS evolution timescale models, 
we have measured the WD age of NGC 2477 to be 1.035 $\pm$ 0.054 $\pm$ 0.087 
Gyr (uncertainties represent the goodness of model fits and discrepancy among 
models, respectively).  This age is consistent with that measured from 
traditional isochrone fitting (both from MSTO fitting by Kassis et al. 1997, 
and the initial WD age derived by von Hippel et al. 1995), as we would expect, 
but our new result is objective, marginalizes over the other cluster 
parameters, and includes posterior distributions on all parameters.  Even with 
only a few WDs, our technique achieves a 5\% age precision within a given 
model, and when marginalizing over three MS stellar evolution models, it achieves 
better than 9\% precision.

\begin{acknowledgements}

   This material is based upon work supported by the National Aeronautics and
Space Administration under Grant  NAG5-13070 issued through the Office of
Space Science, and by the National Science Foundation through Grant
AST-0307315, as well as NSF grant DMS-0907522.

\end{acknowledgements}


\vspace{0.2in}
\begin{center}\textbf{REFERENCES}\end{center}

\noindent
Baggett, S. et al. 2002, HST WFPC2 Data Handbook v. 4.0, ed. B. Mobasher, Baltimore, STScI \\
Bedin, Luigi R., Salaris, Maurizio, Piotto, Giampaolo, King, Ivan R., Anderson, Jay, Cassisi, Santi, Momany, Yazan 2005, ApJ, 624, L45 \\
Bedin, L. R., Salaris, M., Piotto, G., Cassisi, S., Milone, A. P., Anderson, J., King, I. R.  2008a, ApJ, 679, L29 \\
Bedin, L.R., King, I.R., Anderson, J., Piotto, G., Salaris, M., Cassisi, S., \&  Serenelli, A. 2008b, ApJ, 678, 1279 \\
Bellazzini, M., Ibata, R., Monaco, L., Martin, N., Irwin, M. J., \& Lewis, G.F. 2004, MNRAS, 354, 1263 \\
Bellini, A. et al. 2010, A\&A, 513, 50 \\
Bergeron, P., Wesemael, F., \& Beauchamp, A. 1995, PASP, 107, 1047 \\
Bertin, E. \& Armouts, S. 1996, A\&AS, 117, 393 \\
Carraro, G., Vázquez, R.A., Moitinho, A., \& Baume, G. 2005, ApJ, 630, L153 \\
Chaboyer, B., Demarque, P., \& Sarajedini, A. 1996, \apj, 459, 558 \\
Claver, C. F. 1995, PhD. Thesis, The University of Texas at Austin \\
Claver, C. F., Liebert, J., Bergeron, P. 2001, ApJ, 563, 987 \\
DeGennaro, S., von Hippel, T., Jefferys, W.H., Stein, N., van Dyk, D., \& Jeffery, E. 2009, ApJ, 696, 12 \\
Deloye, C. \& Bildsten, L. 2002, ApJ, 580, 1077 \\
Dias, W.S., Assafin, M., Flório, V., Alessi, B.S., \& Líbero, V. 2006, A\&A, 446, 949 \\
Dolphin, A.E. 2000, PASP, 112, 1397 \\
Dotter, A., Chaboyer, B., Jevremovic, D., Kostov, V., Baron, E., \& Ferguson, J.W. 2008, ApJS, 178, 89 \\
Eigenbrod, A., Mermilliod, J.-C., Claria, J. J., Andersen, J., Mayor, M. 2004, A\&A,423,189 \\
Friel, E.D. \& Janes, K.A. 1993, A\&A, 267, 75 \\
Garcia-Berro, E. et al. 2009, Nature, 465, 194 \\
Girardi, L., Bressan, A., Bertelli, G., \& Chiosi, C. 2000. A\&AS, 141, 371 \\
Hansen, B.M.S., et al. 2007, ApJ, 671, 380 \\
Hansen, B.M.S. 2005, ApJ, 635, 522 \\
Hansen, B.M.S., et al. 2002, ApJ, 574, L155 \\
Hartwick, F.D.A., Hesser, J.E., McClure, R.D. 1972, ApJ, 174, 557 \\
Holtzmann, J.A., Burrows, C.J., Casertano, S., Hester, J.J., Watson, A.M., \& Worthy,   G.S. 1995, PASP, 107, 1065 \\
Jeffery, E.J., von Hippel, T., Jefferys, W.H., Winget, D.E., Stein, N., DeGennaro, S. 2007, ApJ, 658, 391 \\
Kalirai, J. et al. 2007, ApJ, 671, 748 \\
Kassis, M., Janes, K. A., Friel, E. D., \& Phelps, R. L. 1997, AJ, 113, 1723 \\
Kepler, S.O., Kleinman, S.J., Nitta, A., Koester, D., Castanheira, B., Giovannini, O., Costa, A.F.M., Althaus, L. 2007, MNRAS, 375, 1315 \\
Kilic, M., von Hippel, T., Mullally, F., Reach, W.T., Kuchner, M.J., Winget, D.E., \& Burrows, A. 2006, ApJ, 642, 1051 \\
Knox, R.A., Hawkins, M.R.S., \& Hambly, N.C. 1999, MNRAS, 306, 736 \\
Koekemoer, A.M. et al. 2002, HST Dither Handbook, Version 2.0, Baltimore, STScI \\
Leggett, S.K., Ruiz, M.T., \& Bergeron, P. 1998, ApJ, 497, 294 \\
Momany et al. 2001, A\&A, 379, 436 \\
Mestel, L. 1952, MNRAS, 112, 583 \\
Mighell, K.J., \& Rich, R.M. 1995, AJ, 110, 1649 \\
Miller, G. E., \& Scalo, J. M. 1979, ApJS, 41, 513 \\
Montgomery, M.H., Klumpe, E.W., Winget, D.E., \& Wood, M.A. 1999, ApJ, 525, 482 \\
Oswalt, T.D., Smith, J.A., Wood, M.A., \& Hintzen, P. 1996, \nat, 382, 692 \\
Richards, G.T. et al. 2009, ApJS, 180, 67 \\
Richards, G.T. et al. 2006, AJ, 131, 2766 \\
Richer, H.B. 2010, STScI May Symposium  \\
Salaris, M., Weiss, A., \& Percival, S.M. 2004, A\&A, 414, 163 \\
Salaris, M., García-Berro, E., Hernanz, M., Isern, J., \& Saumon, D. 2000, ApJ, 544, 1036 \\
van Dyk, D. A., DeGennaro, S., Stein, N., Jefferys, W. H., and von Hippel, T. 2009, The Annals of Applied Statistics, 3, 117 \\
van Loon, J.T., Boyer, M.L., \& McDonald, I. 2008, ApJ, 680, L49 \\
von Hippel, T. 2005, ApJ, 622, 565 \\
von Hippel, T. 2001, in ASP Conf. Ser. 245, ed. T. von Hippel et al., (San Fransisco: ASP), 190 \\
von Hippel, T. \& Gilmore, G. 2000, \aj, 120, 1384 \\
von Hippel, T., Gilmore, G., \& Jones, D.H.P. 1995, MNRAS, 273, L39 \\
von Hippel, T., Gilmore, G., Nial, T., Robinson, D., \& Jones, D.H.P. 1996, AJ, 112, 196 \\
von Hippel, T., Jefferys, W.H., Scott, J., Stein, N., Winget, D.E., DeGennaro, S., Dam, A., \& Jeffery, E. 2006, ApJ, 645, 1436 \\
Weidemann, V. 2000, A\&A, 363, 647 \\
Winget, D. E., Hansen, C. J., Liebert, J., van Horn, H. M., Fontaine, G., Nather, R. E., Kepler, S. O., \& Lamb, D. Q. 1987, \apj, 315, L77 \\
Wood, M. A. 1992, ApJ, 386, 539 \\
Yi, S. Demarque, P., Kim, Y.-C., Lee, Y.-W., Ree, C.H., Lejeune, T., \& Barnes, S. 2001, ApJS, 136, 417 \\

\newpage


\begin{table}[h!]
  \begin{center}
  \begin{tabular}{ccccc}

    \hline
 Age (Gyr) & $(m-M)_{V}$ & $A_{V}^{a}$ & $[Fe/H]$ &  Source \\
    \hline


  1.0  & 11.43 & 0.93 & \---  &  1 \\
  1.0  & 11.45 & 0.713& 0.00  &  2 \\
  1.0$^{b}$ & \--- & \--- & \---  &  3 \\
  1.0  & \---  & \--- & -0.14 &  4 \\
  1.3  & 11.60 & 0.93 & -0.05 &  5 \\
  1.5  & 11.48 & 0.868& \---  &  6 \\


   \hline
  \end{tabular}
\\$^{1}$Kassis et al. 1997, $^{2}$Salaris et al. 2004, $^{3}$von Hippel, Gilmore, \& Jones, 1995, $^{4}$Eigenbrod et al. 2004, $^{5}$Friel \& Janes 1993, $^{6}$Hartwick, et al. 1972; $^{a}$Average value; $^{b}$White dwarf age

 \end{center}
   \caption[Cluster parameters from the literature for NGC 2477]{Cluster parameters from the literature for NGC 2477.}
      \label{lit2477}
\end{table}


\begin{table}[h!]
  \begin{center}
  \begin{tabular}{ccc}

    \hline
 Field & RA(2000) & Dec(2000)  \\
    \hline


 1 & 07:52:16.81 & -38$^{\circ}$35'40''\\
 2 & 07:52:26.20 & -38$^{\circ}$34'40''\\


   \hline
   \end{tabular}
 \end{center}
   \caption{Coordinates of observed fields for NGC 2477 with
        HST/WFPC2.}
      \label{wfpc2_coords_table}
\end{table}


\begin{table}[h!tb]
  \begin{center}
  \begin{tabular}{ccccccc}

    \hline
 \multicolumn{1}{c}{Date}     &        & Exposure &       \\
 \multicolumn{1}{c}{Observed} & Filter & Time (s) & Field & Cycle \\
    \hline


 1994 Mar 18  & F555W & 8$\times$400  & 1 & 4 \\
 1994 Mar 18  & F814W & 9$\times$400  & 1 & 4 \\
 1996 Mar 25  & F555W & 8$\times$500  & 1 & 6 \\
 1997 Mar 17  & F555W &12$\times$500  & 2 & 6 \\
 1997 Mar 17  & F814W & 8$\times$500  & 2 & 6 \\


   \hline
   \end{tabular}
 \end{center}
   \caption{Summary of observations for NGC 2477 with HST/WFPC2.}
      \label{wfpc2_obs_table2477}
\end{table}


\begin{table}[h]

  \begin{center}
  \begin{tabular}{ccccccc}
                
    \hline
    ID  & $V$ & $\sigma_{V}$ & $V-I$ & $\sigma_{V-I}$ & R.A. & Dec. \\
    \hline
                                                                    
 WD1 & 23.108 & 0.008 & 0.292  & 0.020 & 07:52:22.6 & -38$^{\circ}$35'40.9'' \\
 WD2 & 23.566 & 0.011 & 0.284  & 0.026 & 07:52:28.3 & -38$^{\circ}$34'54.3'' \\
 WD3 & 23.689 & 0.010 & 0.452  & 0.027 & 07:52:22.9 & -38$^{\circ}$35'54.4'' \\
 WD4 & 23.904 & 0.012 & 0.443  & 0.033 & 07:52:12.5 & -38$^{\circ}$36'10.2'' \\
 WD5 & 23.963 & 0.015 & 0.387  & 0.034 & 07:52:25.6 & -38$^{\circ}$35'21.7'' \\
 WD6 & 23.957 & 0.015 & 0.371  & 0.034 & 07:52:23.3 & -38$^{\circ}$34'32.1'' \\
 WD7 & 23.972 & 0.016 & 0.334  & 0.038 & 07:52:21.8 & -38$^{\circ}$34'46.9'' \\

         
   \hline
  \end{tabular}
 \end{center}
   \caption{Table of WD candidates in NGC 2477, including photometry and 
	coordinate information.  Numbering is consistent with Figure 
	\ref{ngc2477_hst_wds}.  We note that only confirmed stellar objects 
	are listed.}
      \label{ngc2477_wds_table}
\end{table}


\begin{table}[h]
  \begin{center}
  \begin{tabular}{lcc}

    \hline
\multicolumn{1}{c}{Cluster}   &   Mean     & \\
\multicolumn{1}{c}{Parameter} &  Value     & $\sigma$\\
    \hline


  $[Fe/H]$   & -0.10    & 0.30  \\
 $(m-M)_{V}$ & 11.460   & 0.22 \\
  A$_{V}$    & 0.750    & 0.10  \\
\hline

   \end{tabular}
 \end{center}
   \caption{Priors used by the Bayesian algorithm.  Mean values are within 
	literature values for these parameters (see Table \ref{lit2477}) and 
	$\sigma$ values are representative of conservative uncertainty for 
	these quantities.  The mean values of these distributions were used as 
	the starting value for MCMC, as mentioned in Section \ref{priors}.}
      \label{priors2477}
\end{table}


\begin{table}[h]
  \begin{center}
  \begin{tabular}{lcccccc}

    \hline
\multicolumn{1}{c}{Cluster}   & DSED &                 &  YY  &               & Girardi &   \\
\multicolumn{1}{c}{Parameter} & Mean & $\sigma_{DSED}$ & Mean & $\sigma_{YY}$ & Mean & $\sigma_{G}$ \\
    \hline


  Age (Gyr)      & 1.08   & 0.06 & 0.94   & 0.05 & 1.10   & 0.05  \\
  $[Fe/H]$       & -0.098 & 0.05 & -0.246 & 0.07 & -0.338 & 0.07  \\
 $(m-M)_{V}$ & 11.378 & 0.07 & 11.504 & 0.06 & 11.363 & 0.06  \\
  A$_{V}$        & 0.591  & 0.03 & 0.720  & 0.03 & 0.615  & 0.03  \\
\hline

   \end{tabular}

 \end{center}
   \caption{Mean and standard deviation of the posterior distributions of
        cluster parameters of NGC 2477 from WD plus MS photometry.  We note 
        that the small sigma on $A_{V}$ should not be taken to imply that the 
	cluster does not exhibit differential reddening (as was mentioned in 
	Section \ref{intro}); rather, our model does not incorporate 
	differential reddening at this time.  The sigma here is an error on 
	the mean reddening.}
      \label{bayes2477wd_results}
\end{table}

\end{document}